\documentstyle[aps,epsf]{revtex}

\newcounter{abc}
\renewcommand{\theequation}{\arabic{equation}\alph{abc}}

\begin{document}

\title{Critical-point finite-size scaling in the microcanonical ensemble}
\author{A.D. Bruce and N.B. Wilding}
\address{Department of Physics and Astronomy,
The University of Edinburgh\\
Edinburgh, EH9 3JZ, Scotland, United Kingdom} 

\maketitle

\begin{abstract}

We develop a scaling theory for the finite-size critical behavior of the
microcanonical entropy (density of states) of a system with a
critically-divergent heat capacity.  The link between the microcanonical entropy
and the  canonical energy distribution is exploited to  establish the former,
and corroborate its predicted scaling form, in the case of the 3d Ising
universality class. We show that the scaling behavior emerges clearly when one
accounts for the effects of the negative background constant contribution to the
canonical critical specific heat. We show that this same constant plays a
significant role in determining the observed differences between the canonical
and microcanonical specific heats of systems of  finite size, in the critical
region.

PACS numbers: 05.20.Gg, 05.70.Jk, 64.60.Fr  \\

\end{abstract}

\pacs{02.70.Lq, 05.70.Ce}

\newpage

\section{Introduction}
Statistical mechanics can be formulated in any of a set of ensembles
distinguished by the relationship between the system and its environment
\cite{hill1}.
The principal members of this set are the microcanonical
(prescribed energy) and canonical (prescribed temperature) ensembles.
In the thermodynamic limit (when it exists) the ensembles yield the same
predictions (and are, in this sense, equivalent) 
and the choice of ensemble is a matter of practical convenience.
The canonical ensemble tends to win this contest because it circumnavigates the
hard-constant-energy constraint imposed by  the microcanonical ensemble.

The two ensembles are, however, not always equivalent \cite{grossrev}. They differ for systems
which are `small' in some sense: inherently small systems such as nuclei
or clusters \cite{grossart}; systems with unscreened long-range forces 
\cite{gravforce} where the thermodynamic limit is problematic; and systems at
critical points \cite{whogoeshere}, which are our  principal 
concern here.

Theoretical studies of critical phenomena are almost invariably conducted within
the framework of the canonical ensemble \cite{mefrenorm}. In consequence there
is no substantive framework within which to interpret computational studies of
microcanonical critical behavior. Such studies do, nevertheless, exist, having been motivated, 
variously, by the belief that the microcanonical framework may have some
computational advantages \cite{creutz} and by the discovery \cite{huller} that,
apparently,  critical anomalies in the microcanonical heat capacity are
significantly enhanced with respect to their canonical counterparts.

This paper goes some way towards supplying the missing framework. We develop
(section 2) a finite-size-scaling theory \cite{fssrev} for the microcanonical
entropy (the density of states) of a system with a critically-divergent heat
capacity. In so doing we have, of necessity, to consider more general questions
about the structure of the density of states of a finite-size system --in
particular the implications  of well-established results for the finite-size
structure of the canonical partition function \cite{borgskot}. 

Though somewhat more than a  phenomenology, our theory falls short of being 
microscopically explicit: to determine an explicit form for the relevant scaling
function we need to appeal (section 3) to Monte Carlo (MC) measurements of the
critical {\em canonical} energy probability distribution (pdf). 

The canonical energy pdf itself has a near-critical finite-size-scaling form
which has featured in a number of studies of critical points in fluids
\cite{adbnbw} and lattice  gauge theories \cite{lgt}. Since energy
fluctuations (like the  critical anomaly in the canonical specific heat which
they control) are relatively weak (by comparison with the fluctuations of the
order parameter, and the divergence of its response function) the degree of 
`scaling' reported in previously measured energy pdfs has been
relatively poor --unsatisfactorily so for our purposes here.
This problem is addressed in section 3. We show
that one can fold out (from the measured
distributions) the sub-dominant (but significant)  non-scaling effects that are associated with the
constant background contribution to the canonical  heat capacity, {\em negative}
in the case of the 3d Ising model \cite{liufish}. This
procedure exposes the underlying behavior,  which manifests scaling to an
impressive degree. In addition to providing us with the platform needed for this
work, this procedure may offer the basis for 
improving the mixed-scaling-field
theory \cite{adbnbw}
of critical points in systems that belong to the Ising universality class
but which do not have full Ising symmetry; recent studies \cite{lgt}
have suggested that the current framework is not fully satisfactory.

The scaling form for the critical energy pdf allows us to determine the scaling
form of the microcanonical entropy. In  section 4 we explore this form and show
that it is consistent with predictions
for  both the bulk-critical limit (as regards
the parameters characterizing the specific heat singularity 
\cite{liufish}) and the
finite-size critical limit (the Fisher-Privman constant
\cite{privfish}). 

The microcanonical entropy also provides us with a unified basis for dealing with
both the canonical and the microcanonical specific heats (section 5). 
We 
show that the `corrections' to the scaling behavior of the canonical specific
heat (the {\em negative} background constant)
have subtle consequences for the microcanonical behavior.
In particular they serve to {\em amplify} the difference between microcanonical and
canonical behavior, and are at least partially responsible for the strength of
the anomaly observed in some  microcanonical studies \cite{huller}.

\subsection{The microcanonical scaling ansatz}

We consider a d-dimensional many-body system of linear dimension $L$; 
we assume hypercubic geometry with periodic boundary conditions. The canonical
partition function is, in principle, a discrete sum over system microstates ($r$)
or system energy levels ($s$):
\setcounter{abc}{1}
\begin{equation}
\label{eq:canonicalpfdisc}
Z(\beta , L) = \sum_r  e^{-\beta E_r} =\sum_s \Omega_s e^{-\beta E_s} 
\end{equation}
\addtocounter{abc}{1}
\addtocounter{equation}{-1}
where $\Omega _s$ is the degeneracy of level $s$. We shall suppose that the
system is sufficiently large that the sum over levels can be replaced by an
integral:
\begin{equation}
\label{eq:canonicalpfdef}
Z(\beta , L) = \int d\epsilon \Omega(\epsilon,L) e^{-\beta L^d \epsilon}
\end{equation}
\setcounter{abc}{0}
where $\epsilon\equiv E/L^d$ is the energy density.
The function $\Omega(\epsilon,L)$  is the {\em density of states}; as we have defined
it, it is a true density, having dimensions of inverse energy. We note that
the transition from the discrete representation (Eq.~\ref{eq:canonicalpfdisc}) to its
continuum counterpart (Eq.~\ref{eq:canonicalpfdef}) requires some care: it is
discussed in Appendix A.

Our microcanonical scaling theory comprises a proposal for the form of the
density of states function.  We  formulate it in two stages. Consider first
a regime remote from critical points or lines of phase coexistence. In such a
regime we make
the general finite-size ansatz \cite{convexity}:
\begin{equation}
\label{eq:omegaansatz}
\Omega(\epsilon,L) \simeq \left[ \frac{-L^ds''(\epsilon) }{2\pi}\right] ^{1/2} 
e^{L^d s(\epsilon)}
\end{equation}
The structure proposed for the prefactor makes this a little more than
simply a {\em definition} of the microcanonical entropy density $s(\epsilon)$.
In its support we note, first, that one may readily verify it explicitly
(Appendix B) in the case of
some simple model systems.
Secondly   
we note the implications for the associated
canonical partition function.
Inserting  Eq.~(\ref{eq:omegaansatz}) into Eq.~(\ref{eq:canonicalpfdef}), 
a saddle-point integration gives
\begin{eqnarray}
Z(\beta , L)& =& \left[\frac{L^d}{2\pi}\right]^{1/2} \int d\epsilon \left[-s''(\epsilon)\right]^{1/2}
e^{L^d\left[s(\epsilon) -\beta \epsilon\right]} \nonumber\\
&=& e^{-L^d f(\beta) } \left[ 1 + O (L^{-d}) \right]
\label{eq:canonicalpfresone}
\end{eqnarray}
where
\begin{equation}
\label{eq:fbdef}
f(\beta) \equiv \beta \hat{\epsilon} - s(\hat{\epsilon})
\end{equation}
and $\hat{\epsilon}$ is the solution of
\begin{equation}
\label{eq:hatepsdef}
\beta= s'(\hat{\epsilon})
\end{equation}
Eq.~(\ref{eq:canonicalpfresone}) recovers the prefactor-free form of the canonical
partition function believed to be widely appropriate in regions  (those where
the saddle point integration is to be trusted) remote from
critical points or lines of phase coexistence 
\cite{borgskot}. We note that this form is achieved by virtue of
the prefactor that {\em does} appear in the density of states ansatz (Eq.~\ref{eq:omegaansatz}), which is
just such as to cancel the contributions made by the fluctuations about the
saddle point \cite{grosserror}. 

The argument we have given leaves open the
possibility of power-law corrections to Eq.~(\ref{eq:canonicalpfresone}). It has
long been believed, and more recently established rather generally
\cite{borgskot}, that the corrections to the leading form are actually {\em
exponentially} small in the system size. Since the saddle-point integration
necessarily generates  power-law corrections, one must suppose that there are
compensating  power-law corrections to the ansatz
(Eq.~\ref{eq:omegaansatz}) for the density of states. This conclusion  serves as
a warning  (already suggested by the double appearance of the function $s(\epsilon)$
in Eq.~(\ref{eq:omegaansatz})) that the microcanonical framework faces problems 
which are skirted in the canonical formalism \cite{microprobs}. 

Now, more specifically, consider a system, of the kind  specified above, in the vicinity
of a critical point. We will suppose that the critical point has a divergent
heat capacity;  where we need to be more specific we shall assume it is a member
of the d=3 Ising universality class (or, more specially still, the d=3 Ising model itself).
Within the microcanonical framework the
critical point of such a system is located by a critical value $\epsilon _c$ of the
energy density, sharply-defined in the thermodynamic limit. We are concerned
with the behavior of the microcanonical entropy for energies in the vicinity of
this critical value. To describe this regime we  introduce the  dimensionless
scaling variable \cite{lgsymmetry}
\begin{equation} 
\label{eq:xdef} 
x\equiv a_{\epsilon} L^{1/\nu _\epsilon} (\epsilon -\epsilon _c) 
\end{equation} 

where $a_{\epsilon}$ is an appropriate scale factor  and the index is defined by
\begin{equation} 
\label{eq:nuedef} 
\frac{1}{\nu _e} = \frac{1-\alpha}{\nu}
\end{equation} 
with $\alpha$ the index (assumed positive) characterizing the
heat capacity divergence, and $\nu$ the correlation length index
\cite{gencrit}.
We now reformulate and extend our  basic ansatz (Eq.~\ref{eq:omegaansatz}) 
with the  proposal that, in a region of sufficiently large $L$ and sufficiently
small $\mid \epsilon - \epsilon_c \mid$ \cite{asympequals}
\setcounter{abc}{1}
\begin{equation} 
\label{eq:revomegaansatz} 
\Omega(\epsilon,L) \simeq \left[ \frac{-L^ds''(\epsilon , L ) }{2\pi}\right] ^{1/2} 
e^{L^d s(\epsilon , L )}
\end{equation} 
with
\addtocounter{abc}{1}
\addtocounter{equation}{-1}
\begin{equation} 
\label{eq:sansatz} 
L^ds(\epsilon , L ) \simeq L^d \left [s_c + \beta _c (\epsilon -\epsilon _c )\right] + \tilde{\cal S} (x)
\end{equation} 
\setcounter{abc}{0}
Here $s_c$ is an unimportant constant, $\beta _c$ is the
critical inverse temperature and $\tilde{\cal S} (x)$ is a finite-size-scaling function,
universal given some convention on the scale factor $a_{\epsilon}$, introduced in 
Eq.~(\ref{eq:xdef}). 
The remainder of this paper is devoted to providing
support for this proposal, and exploring the structure
of the microcanonical entropy scaling function which it introduces.

\section{Determining the scaling function}

It should be possible to determine the finite-size-scaling
function $\tilde{\cal S} (x)$ within the renormalization group  framework \cite{brezin}. We have not done
that. Instead we have chosen to learn what we can about this function
from its signatures in MC studies of the {\em canonical} ensemble.

Consider, then, the implications of the scaling form Eq.~(\ref{eq:sansatz})
for the canonical partition function, Eq.~(\ref{eq:canonicalpfdef}). 
We suppose initially (we shall have to refine the supposition, shortly)
that the relevant part of the energy spectrum
 is adequately captured by
Eq.~(\ref{eq:sansatz}). Then
\begin{equation}
\label{eq:canonicalpfrestwo}
Z(\beta , L) \simeq e^{-L^df_0(\beta)} \tilde {\cal Z}(y)
\end{equation}
where
\begin{equation}
\label{eq:f0def}
f_0(\beta) = \beta \epsilon_c -s_c
\end{equation}
and
\begin{equation}
\label{eq:Zydef}
\tilde {\cal Z}(y) =\int dx \sqrt{ \frac{-\tilde{\cal S}''(x)}{2\pi}} e^{-xy +\tilde{\cal S} (x)}
\end{equation}
while 
\begin{equation}
\label{eq:ydef}
y= a_{\epsilon} ^{-1} L^{1/\nu} (\beta -\beta _c)
\end{equation}
provides a scaling measure of the deviation from the critical temperature.
We have made use of the hyperscaling  relation \cite{gencrit}
which links the correlation length index $\nu$ and the heat capacity index
$\alpha$ through
\begin{equation}
\label{eq:hyperscaling}
\frac{1}{\nu} + \frac{1}{\nu _e} = \frac{2-\alpha}{\nu} = d
\end{equation}
The scaling form of the free energy  follows:
\begin{equation}
\label{eq:freeenscal}
F(\beta , L) \equiv -  \ln Z(\beta , L) \simeq  L^d f_0(\beta) -\ln \tilde {\cal Z}(y) \equiv 
\tilde{F}(\beta , L) 
\end{equation}
The canonical energy pdf
\begin{equation}
\label{eq:canendistdef}
P(\epsilon| \beta, L) \equiv Z^{-1}(\beta , L) \Omega(\epsilon,L) e^{-\beta L^d \epsilon} 
\end{equation}
may also be written in scaling form:
\begin{equation}
\label{eq:canendistscalrel}
P(\epsilon| \beta, L) d\epsilon  \equiv P(x| y, L) dx 
\end{equation}
with 
\begin{equation}
\label{eq:canendistscalform}
P(x| y, L) \simeq  \tilde{\cal Z}^{-1}(y) \sqrt{\frac{-\tilde{\cal S}''(x)}{2\pi}} e^{-xy +\tilde{\cal S} (x)}
\equiv  \tilde{P}(x| y) 
\end{equation}
The scaling predictions for the pdf may be tested by examining its cumulants 
\cite{cumulants},
for which the free energy is a generator:
\begin{equation}
\label{eq:cumdef}
\epsilon ^{(n)} (\beta , L ) \equiv (-1)^{n+1}L^{-nd} \frac{\partial ^n F(\beta , L)}{\partial \beta ^n}
\end{equation}
Eq.~(\ref{eq:freeenscal}) then implies that the cumulants have the
scaling form 
\begin{equation}
\label{eq:cumscal}
\epsilon ^{(n)} (\beta , L ) \simeq \left[ a_{\epsilon} L^{1/\nu _\epsilon} \right ]^{-n}  
\tilde{x}^{(n)}(y) + \epsilon _c \delta_{n,1}
\end{equation}
where the scaled cumulants $\tilde{x}^{(n)}(y)$ are universal functions:
\begin{equation}
\label{eq:scalcumscal}
\tilde{x}^{(n)}(y) = (-1)^{n} \frac{\partial ^n \ln \tilde {\cal Z}(y) }{\partial y^n}
\end{equation}

The  canonical mean of the energy density at criticality ($\beta = \beta _c$)
follows as:
\begin{equation}
\label{eq:meanen}
\langle\epsilon \rangle _c \equiv \epsilon ^{(1)} (\beta _c, L) \simeq \epsilon _c 
+ \left[ a_{\epsilon} L^{(1-\alpha)/\nu} \right ]^{-1}
\tilde{x}^{(1)}(y=0)
\end{equation}
MC measurements on the 3d Ising model using
a range of system sizes  (Fig. 1) are fully consistent with this
behavior.

Eq.~(\ref{eq:freeenscal}) implies, likewise, that the 
canonical variance  of the energy density should have the power
law behavior

\begin{equation}
\label{eq:varen}
\langle\epsilon ^2\rangle_c -
\langle\epsilon\rangle_c^2 \equiv \epsilon ^{(2)} (\beta _c, L) 
\simeq a_{\epsilon}^{-2}L^{-d+\alpha/\nu} \tilde{x}^{(2)}(y=0)
\end{equation}

MC measurements (Fig. 2) are only partially consistent with this prediction: the
power law is confirmed, but with an extrapolation whose intercept is far from
zero. This inconsistency  is reflected in the rather limited success  (Fig. 3)
of attempts to collapse the measured energy pdfs for different system sizes on
to a single scaling form. The source of these problems can  be guessed from  the
implications of Eq.~(\ref{eq:varen}) for the canonical specific heat,  which it
mirrors:  the scaling form fails to capture the effects associated with the 
constant background which constitutes the dominant correction to pure scaling
(power-law divergence) of the  canonical specific heat.

There are two ways to rectify this failure. One might extend the theory to
predict the behavior of the (very) finite systems accessible to MC study;
or one might seek  to correct the MC results to expose the true scaling
behavior. We adopt the latter strategy.

Define
\begin{equation}
\label{eq:Fcorsdef}
\Delta f (\beta , L )\equiv L^{-d} \left [F(\beta , L) -\tilde{F}(\beta , L) \right]
\end{equation}
the difference between the true free energy density and its asymptotic scaling
form (see Eq.~(\ref{eq:freeenscal})). We shall ignore the effects
of confluent singularities: they are not the dominant `corrections to scaling' here.
Then $\Delta f (\beta , L )$ is analytic and may be approximated, near $\beta_c$,
 by the expansion
\begin{equation}
\label{eq:Fcorsres}
\Delta f (\beta , L ) \simeq \sum_{n=0}^{\infty} \Delta f^{(n)}_{c} \frac{(\beta -\beta _c)^n}{n!}
\end{equation}
These additional contributions to
the free energy imply additional contributions to the  energy cumulants
(Eq.~\ref{eq:cumdef}): 
\begin{equation}
\label{eq:cumcor}
\Delta \epsilon ^{(n)} (\beta , L ) \equiv (-1)^{n+1}L^{-(n-1)d} \frac{\partial ^n \Delta f (\beta , L )}
{\partial \beta ^n}
\end{equation}
At criticality Eq.~(\ref{eq:cumscal}) must then be modified to read
\begin{eqnarray}
\epsilon ^{(n)} (\beta _c , L)& =& \left[ a_{\epsilon} L^{1/\nu _\epsilon} \right ]^{-n}  
\left [\tilde{x}^{(n)}(y=0)
+ \Delta x_c ^{(n)}(L) \right ] + \epsilon _c \delta_{n,1} \nonumber\\
&\equiv &
\left[ a_{\epsilon} L^{1/\nu _\epsilon} \right ]^{-n} x^{(n)}(y=0,L)+ \epsilon _c \delta_{n,1}
\label{eq:cumscalcora}
\end{eqnarray}
where 
\begin{equation}
\label{eq:cumscalcorb}
\Delta x_c ^{(n)}(L) \equiv x^{(n)}(y=0,L)- \tilde{x}^{(n)}(y=0)
=(-1)^{n+1} a_{\epsilon} ^n L^{d-n/\nu} \Delta f^{(n)}_{c}
\end{equation}
The $n=1$ correction is absent by {\it fiat}: the choice of $\epsilon _c$
ensures this. The $n \ge 3$ corrections are sufficiently strongly `irrelevant' (they vanish
sufficiently strongly with $L$)  that they may reasonably be neglected.
But the $n=2$ correction decays only slowly:
\begin{equation}
\label{eq:varcor}
\Delta x_c ^{(2)}(L) =-a_{\epsilon} ^2 L^{-\alpha/\nu} \Delta f_c^{(2)}
= a_{\epsilon} ^2 L^{-\alpha/\nu} c_{0c}
\equiv -g(L)
\end{equation}
where the last step {\em defines} $g(L)$ (a convenient parameter)  while
\begin{equation}
\label{eq:cocdef} 
c_{0c} \equiv - \Delta f_c^{(2)} = - \frac{\partial^2 \Delta f (\beta , L )}{\partial \beta ^2}\mid_{\beta_c}
\end{equation}
is identifiable as the constant `background' to the near-critical 
canonical specific heat. With this addition, Eq.~(\ref{eq:varen}) is modified to read
\begin{equation}
\label{eq:varencor}
L^{d}\left[\langle\epsilon ^2\rangle_c -
\langle\epsilon\rangle_c^2 \right] 
\simeq L^{\alpha/\nu}a_{\epsilon} ^{-2} x^{(2)}(y=0,L) = 
L^{\alpha/\nu}a_{\epsilon} ^{-2} \tilde{x}^{(2)}(y=0) + c_{0c}
\end{equation}
which is now fully consistent with the  MC measurements of 
Fig. 2, with (it is to be noted) a {\em negative} value
for $c_{0c}$ \cite{negative}. From a thermodynamic point of view these results
simply reflect the fact that, for any system size practically accessible
to computer-simulation, the `critical'
contribution to the canonical specific heat is not large enough to
dominate the `non-critical background'. But the argument also
shows us how to eliminate the effects of this `background' from the energy pdf.

Consider the cumulant representation \cite{cumulants} of the {\em scaling}
energy pdf (Eq.~\ref{eq:canendistscalform}) at criticality: 
\begin{equation}
\label{eq:Ptxycumrep}
\tilde{P}(x| y=0) =\frac{1}{2\pi} \int_{-\infty}^{\infty} d \tau e^{ix\tau}
\exp \left[ \sum_{n=1}^{\infty} \frac{(-i\tau)^n}{n!} \tilde{x}^{n}(y=0) \right]
\end{equation}
The corresponding relation for the {\em observed} energy pdf at criticality, written in its inverse form,
is
\begin{equation}
\label{eq:Pxycumrep}
\exp \left[ \sum_{n=1}^{\infty} \frac{(-i\tau)^n}{n!} x^{(n)}(y=0,L) \right]
= \int_{-\infty}^{\infty} d x' e^{-ix'\tau} 
P(x'| y=0, L)
\end{equation}
Appealing to the our conclusion
that, for large enough $L$,  the cumulants of the two pdfs differ significantly only in 
the $n=2$ case, and using  Eqs. (\ref{eq:cumscalcorb}) and (\ref{eq:varcor}), we find that
\begin{eqnarray}
\tilde{P}(x| y=0)& =&
\frac{1}{2\pi} 
\int_{-\infty}^{\infty} d \tau 
\int_{-\infty}^{\infty} d x' 
e^{i(x-x')\tau -g(L) \tau^2/2}
P(x'| y=0, L) \nonumber
\\
&=&
\frac{1}{\sqrt{2\pi g(L)}} 
\int_{-\infty}^{\infty} d x' 
e^{-(x-x')^2/[2g(L)]}
P(x'| y=0, L)
\label{eq:PtxyPxy}
\end{eqnarray} 

This result shows that the {\em scaling} form of the critical  pdf may be
exposed by {\em convolution} of the {\em observed} (and thus, generally,
non-scaling) pdfs with gaussians whose widths are controlled by the  specific
heat background. Note that the argument rests on the fact that this background
is negative (so that $g(L)$ as defined in Eq.~(\ref{eq:varcor}) is positive). If
the background constant were positive  our argument would have to be
restructured to prescribe the   scaling form by a process of  {\em
de}convolution, which is  numerically problematic. As it is, the convolution
process can be implemented easily. With $c_{0c}$ fixed by the ordinate intercept
in Fig. 2, the pdfs measured on different system sizes can each be corrected in
this way to yield estimates of the scaling pdf. The results are shown in Fig. 4.
The improvement with respect to the raw data (Fig. 3) is striking. This
improvement reflects not only the removal of the non-scaling contribution to the
second cumulant but also that the requisite convolution process provides a {\em
natural} smoothing of the MC data \cite{naturalsmooth}. The consequences of this
correction for the {\em shape} of the distribution are also striking. The
skewness \cite{skewness} clearly visible in the raw distributions (Fig. 3) is
largely suppressed to expose a scaling form that is, at first appearance,
gaussian.  Indeed the portion of the distribution  evident on the  scale of Fig.
4 is  gaussian  to within deviations  of a few per cent. However  the behavior
in the wings (evident on the logarithmic scale utilized in Fig. 5) is markedly
different on the high- and low-energy sides. 

The scaling of the critical energy pdf corroborates the scaling of the
microcanonical entropy (cf Eq.~(\ref{eq:canendistscalform})). Given the double appearance
of $\tilde{\cal S} (x)$ in Eq.~(\ref{eq:canendistscalform})
it is practical to infer only the `effective'
microcanonical entropy scaling function 
\begin{equation}
\label{eq:effSdef}
\tilde{\cal S}_{eff} (x) \equiv \tilde{\cal S} (x) + \frac{1}{2}\ln\left[{\frac{-\tilde{\cal S}''(x)}{2\pi}}\right]
\end{equation}
Fig. 5 shows the form implied  by Eq.~(\ref{eq:canendistscalform})
\begin{equation}
\tilde{\cal S}_{eff} (x) = 
\tilde{\cal S}_{eff} (x=0) 
+\ln \left[ \frac{\tilde{P}(x| y=0)}
{\tilde{P}(x=0| y=0)} \right]
\label{eq:effScalc}
\end{equation}
We note as a matter of empirical fact that
$\tilde{\cal S}_{eff} (x)$ is concave. The concavity of $\tilde{\cal S} (x)$ itself is already assumed in
our basic scaling ansatz \cite{convexity}.

\section{The scaling theory: implications and tests}

Although we have no first-principles calculation of
the scaling function $\tilde{\cal S} (x)$ to offer here we can identify, and test, some of the
properties it must have to match anticipated behavior in both the
thermodynamic and finite-size-critical limits.

We consider, in particular, the limiting large $\mid x \mid$ behavior. 
In this regime we anticipate that
\begin{equation}
\label{eq:largex}
\tilde{\cal S} (x) \simeq -b_{\pm} \mid x \mid ^{\theta} + r_{\pm} 
\hspace*{0.5cm}
( \mid x \mid \gg 1)
\end{equation}
where the $+$ and $-$ subscripts refer, respectively, 
to the regions of positive and negative $x$. To make explicit
identifications of the new quantities introduced in this equation (the exponent
$\theta$ and the amplitudes $b_{\pm}$, $r_{\pm}$), consider the
scaling part of the partition function (Eq.~\ref{eq:Zydef}). In the limit of
large $\mid y \mid$ the integral in Eq.~(\ref{eq:Zydef}) is dominated by 
one or other of the large $\mid x \mid$ regimes. Substituting Eq.~(\ref{eq:largex}),
a saddle-point integration yields
\begin{equation}
\label{eq:largey}
\ln \tilde {\cal Z}(y) \simeq a_{\pm} \mid y \mid ^{\theta/(\theta-1)} + r_{\pm} 
\hspace*{0.5cm}
( \mid y \mid \gg 1)
\end{equation}
where the $+$ and $-$ subscripts now refer, respectively, to the regions of {\em
negative} and {\em positive} $y$ \cite{whyplusminus}, and 
\begin{equation}
\label{eq:abratrel}
\frac{a_+}{a_-} = \left(\frac{b_-}{b_+} \right ) ^{1/(\theta -1)}
\end{equation}
As in the argument leading to Eq.~(\ref{eq:canonicalpfresone}) the fluctuations
about the saddle point are canceled by the pre-exponential factor in Eq.~(\ref{eq:Zydef}) to leave 
power-law (`ln-free') behavior \cite{prefactor}.

The thermodynamic limit of the near-critical free energy, defined by Eq.~(\ref{eq:freeenscal}), now follows
as
\begin{equation}
\label{eq:freeentdlim}
\tilde{F}(\beta , L) \simeq  L^d\left[ f_0(\beta) - A_{\pm} \mid \beta -\beta _c \mid ^{2-\alpha} \right ] -r_{\pm} 
\end{equation}
where we have identified 
\begin{equation}
\label{eq:thetadef}
\theta = \frac{2-\alpha}{1-\alpha}
\end{equation}
and (given Eq.~\ref {eq:abratrel})
\begin{equation}
\label{eq:aratdef}
\frac{A_+}{A_-} 
=\frac{a_+}{a_-} 
= \left(\frac{b_-}{b_+} \right ) ^{1-\alpha}
\end{equation}

To establish the role of the remaining constants ($r_{\pm}$) 
in Eq.~(\ref{eq:largex}) we consider the anomalous 
contribution to the free energy \cite{privfish} defined by
\begin{equation}
F_a(\beta)\equiv \lim _{L\rightarrow \infty} \left\{\tilde{F}(\beta , L) - L^d 
\lim _{L\rightarrow \infty} \frac{\tilde{F}(\beta , L)}{L^d} \right\}
\label{eq:anomalydef}
\end{equation}
Appealing to Eq.~(\ref{eq:freeentdlim}), and recalling our sign convention
\cite{whyplusminus}, we identify
\begin{equation}
\label{eq:anomalypmone}
F_a(\beta) = \left\{ \begin{array}{ll} 
- r_+ & (\beta < \beta _c)\\
- r_- & (\beta >\beta _c)
\end{array}
\right .
\end{equation}
On the basis of rather general arguments 
\cite{borgskot}
we expect that {\em away} from a critical point the free energy anomaly is
just minus the logarithm of the number of coexisting phases, so that
\setcounter{abc}{1}
\begin{eqnarray}
r_+ &=&  -F_a(\beta <\beta _c) = 0 \label{eq:anomalypmtwoa}\\
\addtocounter{abc}{1}
\addtocounter{equation}{-1}
r_- &=&  -F_a(\beta >\beta _c) = \ln 2
\label{eq:anomalypmtwob}
\end{eqnarray}
\setcounter{abc}{0}
In the critical finite-size limit we find from
Eq.~(\ref{eq:freeenscal}) 
\begin{equation}
\label{eq:freeencritlim}
\tilde{F}(\beta_c,L)= L^df_0(\beta_c) - \ln \tilde {\cal Z}(0)
\end{equation}
The {\em critical} value of the free energy anomaly, defining the Privman
-Fisher constant $U_0$ \cite{privfish}, follows as
\begin{equation}
\label{eq:anomalycritval}
U_0 \equiv F_a(\beta_c) = - \ln \tilde {\cal Z}(0)
\end{equation}

These predictions are testable to varying degrees through both the
energy-dependence of the energy pdf and the temperature-dependence of the
associated free energy.

Figure 6 shows the results for the  ratio of the specific heat amplitudes that
follow from Eq.~(\ref{eq:aratdef}) when the measured decay of the critical energy
pdf (Figs. 4, 5) is matched to the prediction (\ref{eq:largex}), in conjunction
with Eq.~(\ref{eq:canendistscalform}). We can expect  the predictions and
observations to match up only in a window of $x$ values. Clearly, $x$ must be
large enough to lie within the thermodynamic critical region; but it must also
not be so large that the associated energy lies outside the domain of validity
of the basic  scaling ansatz (Eq.~\ref{eq:sansatz}). The {\em size} of this
window should increase with increasing system size. The {\em location} of this
window on the $x$-axis may also be expected to be different for the positive and
negative $x$-branches --if, as seems reasonable, one regards the {\em
correlation length} $\xi$ (rather than $x\sim \epsilon -\epsilon_c$ or $y \sim \beta
-\beta_c$) as a measure of  criticality.  This is, indeed, the view we have
adopted \cite{alsoliufish}. Thus  Fig. 6 shows the results for the `effective' amplitude ratio, 
obtained by fitting over ranges of $x$ values, with each pair of (positive and
negative ranges) being centered on the same value of $z=L/\xi$, used as the
abscissa \cite{findingxi}. On the basis of this  data \cite{givenalpha} 
we make  the assignment
$A_+/A_- = 0.575(10)$ which  is to be compared with $A_+/A_- = 0.523 (9)$ in
reference \cite{liufish} and $0.567 (16)$ in reference \cite{hasenbusch}.
 
In Fig. 7 we show the results for   $\ln \tilde {\cal Z}(y)$ that follow  (cf
Eqs. (\ref{eq:Zydef}), (\ref{eq:effSdef}), (\ref{eq:effScalc})) from the measured
energy pdfs, using Eq.~(\ref{eq:effScalc}). The latter determines $\tilde{\cal S}_{eff} (x)$ only to
within an additive constant which must be fixed by appeal to the predicted value
of either $r_+$ or $r_-$. We have chosen the latter so that
Eq.~(\ref{eq:anomalypmtwob}) is satisfied, by {\it fiat}.
The motivation for this choice is that it provides us with an inherently more
reliable estimate of the parameter
$U_0$ (which, unlike $r_{\pm}$, is not known a priori). Since $-U_0=\ln \tilde {\cal Z}(0)$
is closer to $r_-$ than to $r_+$ the function $\ln \tilde {\cal Z}(y)$ converges more quickly
to its $y>0$ asymptote than it does to its $y<0$ asymptote. Fixing $r_-$
(the intercept of the $y>0$ asymptote) thus tethers the value assigned to $U_0$ 
more effectively than fixing $r_+$.     
As with the amplitude ratio considered above,
the value assigned to $U_0$ 
depends upon the range of $y$-values  used in the fit to the
anticipated asymptotic form (Eq.~\ref{eq:largey}). Again we have chosen to
characterize the temperature range utilized  through the 
value of the ratio $z=L/\xi$; again we can
expect the analysis to be trustworthy only if it is based upon data lying
within the thermodynamic-critical window.
Our data (Fig. 8) do not allow 
a systematic analysis of the approach to the desired limit;  but they provide
the basis for the assignment
$U_0=-0.57 (2)$ \cite{givenalpha}. The assignment of the uncertainty limit is
subjective but, we think, conservative. We note the close correspondence 
with the assignment ($U_0=-0.57$) emerging from an earlier study
\cite{adb}, similar in concept, but utilizing the distribution of the order
parameter. However our assignment differs (in what would seem to be a statistically
significant fashion) from
the result $U_0=-0.625 (5)$ obtained by Mon \cite{mon} on the basis of
altogether different techniques.

\section{Microcanonical and canonical specific heats}

\subsection{Generalities}

Thus far we have focused on the implications of the {\em microcanonical} entropy
for observations made in the {\em canonical} ensemble. We now turn to
consider their implications for observations made within ensembles that are (or
are approximations to) microcanonical.

We will assume (in keeping with eg \cite{grossrev,huller})
that the temperature of a 
microcanonical system should be identified from the relation
\begin{equation}
\label{eq:microbetadef}
\beta^{\mu e}(\epsilon , L)  =L^{-d} \frac{\partial\ln \Omega(\epsilon,L)}{\partial \epsilon}
\end{equation}
This identification is certainly required in the thermodynamic limit; but in the
context of systems of finite size it is, it seems, a matter of convention
\cite{microbetadef}. 

It is illuminating to link this temperature with {\em canonical} observables.
Appealing to Eq.~(\ref{eq:canendistdef}) we may write
\begin{equation}
\label{eq:microbeta}
\beta^{\mu e}(\epsilon , L) = \beta + L^{-d} \frac{\partial \ln P(\epsilon| \beta, L)}{\partial \epsilon} 
\end{equation}
where (notwithstanding appearances to the contrary) 
the rhs depends on $\epsilon$ but not $\beta$. This result shows that
the equation prescribing the microcanonical temperature for a given energy
is just the inverse of the equation prescribing the  {\em most probable
energy} for a given temperature:
\begin{equation}
\label{eq:microbetaeps}
\beta = \beta^{\mu e}(\epsilon , L) 
\hspace*{0.5cm}
\Longleftrightarrow
\hspace*{0.5cm}
\epsilon = \hat{\epsilon}^{ce} (\beta, L)
\end{equation}
By comparison, within the canonical ensemble itself, `the'
energy  for a given temperature is customarily identified with the canonical
{\em mean}:
\begin{equation}
\label{eq:canbetaeps}
\epsilon = \bar{\epsilon}^{ce} (\beta, L)
\end{equation} 
Eqs. (\ref{eq:microbetaeps}) and (\ref{eq:canbetaeps}) make it immediately plain 
that the energy-temperature relationships associated 
with the two ensembles will coincide 
to the extent that the canonical energy distribution is {\em gaussian}
(and thus  has coincident mean $\bar{\epsilon}^{ce}$ and mode $\hat{\epsilon}^{ce}$).
This correspondence is guaranteed in the thermodynamic limit, but not (in
general) when finite-size effects are significant.

The energy-temperature relationships  are most usually probed through their
derivatives, the associated specific heats.
In the microcanonical case
\begin{equation}
\label{eq:microsh}
c^{\mu e}(\epsilon , L) = 
- \left[ \frac{\partial  \beta^{\mu e}(\epsilon , L)}{\partial \epsilon}\right]^{-1} =
- L^{d} \left[\frac{\partial^2 \ln P(\epsilon| \beta, L)}{\partial \epsilon^2} \right]^{-1}
\end{equation}
where, again, the $\beta$-dependence of the rhs is illusory.

In the canonical case (appealing to Eq.~(\ref{eq:cumdef}))
\begin{equation}
\label{eq:cansh}
c^{ce}(\beta , L) = -  \frac{\partial \bar{\epsilon}^{ce} (\beta, L) }{\partial \beta} 
= L^{d} \epsilon ^{(2)} (\beta , L)
\end{equation}

Like the two `caloric equations of state'
(Eqs. (\ref{eq:microbetaeps}) and (\ref{eq:canbetaeps})) these two specific heats
are guaranteed to agree in the thermodynamic limit; but they differ (in general)
in the finite-size critical regime to which we now turn.

\subsection{Scaling forms}

First, we examine
the asymptotic scaling regime where the the `background' contribution to the
specific heat  can be neglected. We will consider the consequences of the
corrections associated with the latter in the following section.

In the scaling regime where the canonical energy pdf 
may be represented by its scaling form (Eqs.~(\ref{eq:canendistscalrel})
and (\ref{eq:canendistscalform})) Eq.~(\ref{eq:microbeta}) can be rewritten in
terms of the energy and temperature scaling  variables
(Eqs.~(\ref{eq:xdef}) and (\ref{eq:ydef})) as
\begin{equation}
\label{eq:microbetascal}
y^{\mu e}(x) = y +  \frac{\partial \ln \tilde{P}(x| y)}{\partial x}  = 
\frac{\partial \tilde{\cal S}_{eff} (x)} {\partial x} 
\end{equation}
where in the last step 
we have exercised the right to set $y=0$ 
(the result is independent of $y$) and have made use of
Eqs. (\ref{eq:canendistscalform}) and (\ref{eq:effSdef}).
The microcanonical specific heat (Eq.~\ref{eq:microsh}) follows in scaling form
\setcounter{abc}{1}
\begin{equation}
\label{eq:microshscala}
c^{\mu e}(\epsilon , L) \simeq L^{\alpha/\nu} a_{\epsilon}^{-2} 
\tilde{c}^{\mu e}(x)
\end{equation}
\addtocounter{abc}{1}
\addtocounter{equation}{-1}
with
\begin{equation}
\tilde{c}^{\mu e}(x) =  -  \left[\frac{\partial^2 \ln \tilde{P}(x| y=0)}{\partial x^2} \right]^{-1}
= -\left[ \frac{\partial^2 \tilde{\cal S}_{eff} (x) }{\partial x^2} \right]^{-1}
\label{eq:microshscalb}
\end{equation}
\setcounter{abc}{0}

The scaling form of the canonical specific heat follows in a similar fashion,
using Eqs. (\ref{eq:cumscal}) and
(\ref{eq:cansh})
\setcounter{abc}{1}
\begin{equation}
\label{eq:canshscala}
c^{ce}(\beta , L) \simeq L^{\alpha/\nu} a_{\epsilon}^{-2} \tilde{c}^{ce}(y)
\end{equation}
\addtocounter{abc}{1}
\addtocounter{equation}{-1}
with (Eq.~\ref{eq:varen})
\begin{equation}
\label{eq:canshscalb}
\tilde{c}^{ce}(y) =  \tilde{x}^{(2)}(y)
\end{equation}
\setcounter{abc}{0}

The forms of both the scaling functions $\tilde{c}^{\mu e}(x)$ and $\tilde{c}^{ce}(y)$ can be determined
from the scaling form for the microcanonical entropy (Fig. 5) or,
equivalently, the critical canonical energy pdf (Fig. 4),
established  in the preceding section. They are compared in
Fig. 9.  In the  microcanonical case we have used
Eq.~(\ref{eq:microbetascal}) to identify the microcanonical temperature 
$y=y^{\mu e}(x)$ to be associated with a given value of the energy variable $x$. 

In the thermodynamic limit realised at  large values of $\mid y \mid$ the
two functions are, necessarily,  consistent with one another, and approach the asymptotic behavior implied by
Eq.~(\ref{eq:freeentdlim}).
In the finite-size-critical (small $\mid y \mid$) regime,  however, 
clear differences between the two scaling forms are apparent. In particular,
the microcanonical maximum exceeds the canonical maximum by some $10\%$.
One can show (Appendix C) that this --the fact that the microcanonical maximum is
the larger one-- follows necessarily if the
scaling function $\tilde{\cal S} (x)$ is concave.

The two scaling functions cross very close to the point ($y=0$) identifying the
bulk critical temperature. One can see this already in published microcanonical
data \cite{huller}; a similar  `ensemble-independence' has also been noted in
studies of the  `gaussian ensemble' \cite{gaussensemble}. We have been unable to
see any deep reason for this correspondence; but we do not discount the
possibility that there is one.

Though clearly visible, the differences between the two scaling functions
are smaller than suggested by existing MC data \cite{huller}. 
The next section explains why.

\subsection{Beyond scaling: the role of the `background'}

To understand the behavior observed in MC studies of microcanonical behavior,
we must allow for the `corrections to scaling ' which, in the canonical ensemble,
are reflected  simply in the existence of the additive negative background contribution
to the heat capacity; their signature in the microcanonical  ensemble is more
subtle.

The differences between the canonical and microcanonical results is effectively
a strongly anharmonic effect: in a system of finite size, critical fluctuations
sample a region of the entropy surface sufficiently large that the
variation of its curvature becomes significant. We can expose the consequences
analytically within an anharmonic perturbation theory in the cumulants of the
energy pdf.  The calculation is straightforward and we describe it in
outline only.

We appeal to the cumulant representation of the energy pdf at some (general)
temperature:
\begin{equation}
\label{eq:Pgencumrep}
P(\epsilon| \beta, L) =\frac{1}{2\pi} \int_{-\infty}^{\infty} d \tau e^{i\epsilon\tau}
\exp \left[ \sum_{n=1}^{\infty} \frac{(-i\tau)^n}{n!}\epsilon ^{(n)} (\beta , L ) \right]
\end{equation}
We expand perturbatively to first order in the fourth cumulant and to second order
in the third. 
We evaluate the second derivative of the logarithm of this function, which
determines (cf Eq.~(\ref{eq:microsh})) 
the  microcanonical specific heat associated with a given energy density.
We evaluate this function at the modal energy $\hat{\epsilon}=\hat{\epsilon}(\beta, L) $ associated with the chosen
temperature, prescribed  by the (perturbative) solution of the microcanonical
caloric equation of state (Eq.~\ref{eq:microbeta}).
The result is
\begin{equation}
\label{eq:shpertheory}
c^{\mu e}(\hat{\epsilon},L) 
=c^{ce}(\beta , L) 
\left\{
1- \frac{
\epsilon^{(4)}(\beta, L)
\epsilon^{(2)}(\beta, L)
-\left[\epsilon^{(3)}(\beta, L)\right]^2
}
{2\left[\epsilon^{(2)}(\beta, L)\right]^3}
+ \ldots
\right\}
\end{equation}

Eq.~\ref{eq:cumdef} shows that the cumulant correction terms displayed in
this equation are $O(L^{-d})$ in the thermodynamic limit, confirming the equality of
microcanonical and canonical predictions in this limit. To see what happens in
the finite-size-critical region we focus (for simplicity) on the temperature $\beta_m$ for 
which the canonical
specific heat is maximal, identified by the solution of
\begin{equation}
\label{eq:ccemax}
\epsilon^{(3)}(\beta, L)=0 = \frac{d c^{ce}(\beta , L)}{d \beta} 
\end{equation}
At this temperature Eq.~(\ref{eq:shpertheory}) simplifies to
\begin{equation}
\label{eq:shpertheorybma}
c^{\mu e}(\hat{\epsilon}_m , L) 
=c^{ce}(\beta_m, L) 
- \frac{L^d\epsilon^{(4)}(\beta _m, L)}
{2\epsilon^{(2)}(\beta _m, L)}
+ \ldots
\end{equation}
where $\hat{\epsilon} _m =\hat{\epsilon}(\beta_m, L)$,
and we have used Eq.~(\ref{eq:cansh}).
Now we appeal to the scaling forms for the cumulants (Eq.~\ref{eq:cumscal}),
and fold in the effects of the additional non-scaling contribution to the 
second cumulant (Eq.~\ref{eq:varcor}) to conclude that

\begin{equation}
\label{eq:shpertheorybmb}
c^{\mu e}(\hat{\epsilon}_m , L) 
=c^{ce}(\beta_m, L) 
- \frac{L^{\alpha/\nu} a_{\epsilon} ^{-2}\tilde{x}^{(4)}(y_m)}
{2\left[\tilde{x}^{(2)}(y_m) -g(L)\right]}
+ \ldots
\end{equation}

This result makes clear (albeit perturbatively) that, in the 
finite-size-limited regime, the
temperature-independent additive  `background constant' in the canonical specific heat 
(manifested in the parameter $g(L)$)
does not simply translate into an additive 
energy-independent background in its microcanonical counterpart. 

To expose the implications for the
difference between canonical and microcanonical
specific heats we introduce the dimensionless parameter:
\begin{equation}
\label{eq:Rdef}
R(L) \equiv \frac{c^{\mu e}(\hat{\epsilon}_m , L) - c^{ce}(\beta_m , L) }
{c^{ce}(\beta_m , L)}
= - \frac
{\tilde{x}^{(4)}(y_m)}
{2\left[\tilde{x}^{(2)}(y_m) -g(L)\right]^2}
+ \ldots
\end{equation}
Then
\begin{equation}
\label{eq:Rrat}
\frac{R(L)}{R(\infty)}
=\left[
\frac{\tilde{x}^{(2)}(y_m)}{\tilde{x}^{(2)}(y_m) -g(L)}
\right]^2
=\left[1-
\frac
{c_{0c}}
{c^{ce}(\beta_m , L)}
\right]^2
\end{equation}
where $R(\infty)$ is the scaling limit  of $R(L)$.

The significance of the background constant $c_{0c}$ --in particular,  its {\em
sign}-- is now apparent. The {\em negative} value of this constant results in an {\em
amplification} of the difference between the microcanonical and  canonical results  (at
$\beta_m$),  to a degree that diminishes with increasing system size.  This is
not simply the trivial effect that would arise from a uniform (downward)  shift
of both functions: Eq.~(\ref{eq:shpertheorybmb}) shows that this is not what
happens,  as does the power of two on the rhs of  Eq.~(\ref{eq:Rrat}). 

It is not hard to track down the origins of this effect. The difference between
the canonical and microcanonical specific heats is, we have noted, 
an anharmonic effect; in the present context the `corrections to scaling' 
{\em reduce} (only) the second cumulant of the energy pdf and thus, in a
relative sense, 
{\em enhance} the anharmonic (non-gaussian) character of the energy pdf, as one can
see immediately from a comparison of Figures 3 and 4.

The effect is significant. For $L=10$ (as used in 
the simulations reported in \cite{huller}), estimating 
$c^{ce}(\beta _m , L)$ by $c^{ce}(\beta_c , L)$ one can read off from
Fig. 2 that
$R(L)/R(\infty) \sim 4 $. The somewhat unexpected
conclusion that the fractional difference between 
$c^{ce}$ and $c^{\mu e}$ at bulk criticality actually {\em decreases} for increasing
$L$ is consistent with some MC studies \cite{ray}.

\section{Conclusions}

We review, briefly, the three principal strands of this work.

First, we have broached the general question of  the finite-size
corrections to the density of states of a many-body system. The explicit
proposal for the pre-exponential structure advanced in  Eq.~(\ref{eq:omegaansatz})
is consistent with the pre-factor-free structure of the canonical partition
function  \cite{borgskot} and with the behavior of the simple  models discussed
in Appendix B. Given the growing interest in the behavior of
mesoscopically-sized systems, this proposal seems to merit some further study,
with more rigor than we have attempted to offer here.

Second, we have shown how one can fold out from the  canonical 
energy fluctuation spectrum the principal corrections to scaling. 
The underlying  behavior exhibits scale-invariance to a degree that seems
remarkable, given the relative weakness of energy fluctuations. 
It is, we have seen, also largely consistent with established 3d-Ising critical properties.

Third, we have provided a finite-size scaling theory of the microcanonical
ensemble. This was the original motivation for this work-- specifically,  the
suggestion \cite{huller} that the finite-size-smearing of critical behavior 
characteristic of the canonical ensemble is `greatly reduced' within the
microcanonical framework. Reference~\cite{huller} offers two pieces 
of supporting evidence for this contention, which merit final comment. 

Reference~\cite{huller} suggests, firstly, that, in the vicinity of
$\epsilon_c$,  the microcanonical entropy (measured with the techniques
described in \cite{dynensemble}) can be adequately represented by a form (Eq. 6 of
reference~\cite{huller}) which allows for {\em no} finite-size corrections at
all, and which corresponds essentially to the large $x$ limit
(Eq.~\ref{eq:largex}) of our scaling function. In fact the quality of the fit
provided by this representation is rather poor.
And we would expect it to be so. The
measured microcanonical entropy evolves in a {\em manifestly smooth} way
\cite{straightline} between the limiting thermodynamic forms appropriate above
and below $\epsilon_c$; Eq. 6 of reference~\cite{huller} is non-analytic at
$\epsilon_c$. Moreover, in analyzing data for the entropy and its derivatives,
it is -- we have seen -- essential (on all systems practically accessible) to do
justice to the corrections associated with the background constant $c_{0c}$.
Even in the thermodynamic limit the corrections allowed for in Eq.~(8) of
reference \cite{huller} do not do this.  

The second piece of supporting evidence
offered in reference \cite{huller} is a  striking enhancement of the  critical
peak in the microcanonical specific heat, with respect to its canonical
counterpart. As we have seen,  this behavior is at least partly due
\cite{resolution} to the effects associated with $c_{0c}$; Figure 9 indicates
that the underlying differences are rather less dramatic.

\acknowledgements

NBW acknowledges the financial support of the Royal Society (grant no.
19076), the EPSRC (grant no. GR/L91412) and the Royal Society of
Edinburgh.

\newpage


\begin{figure}[h]
\epsfxsize=160mm \epsffile{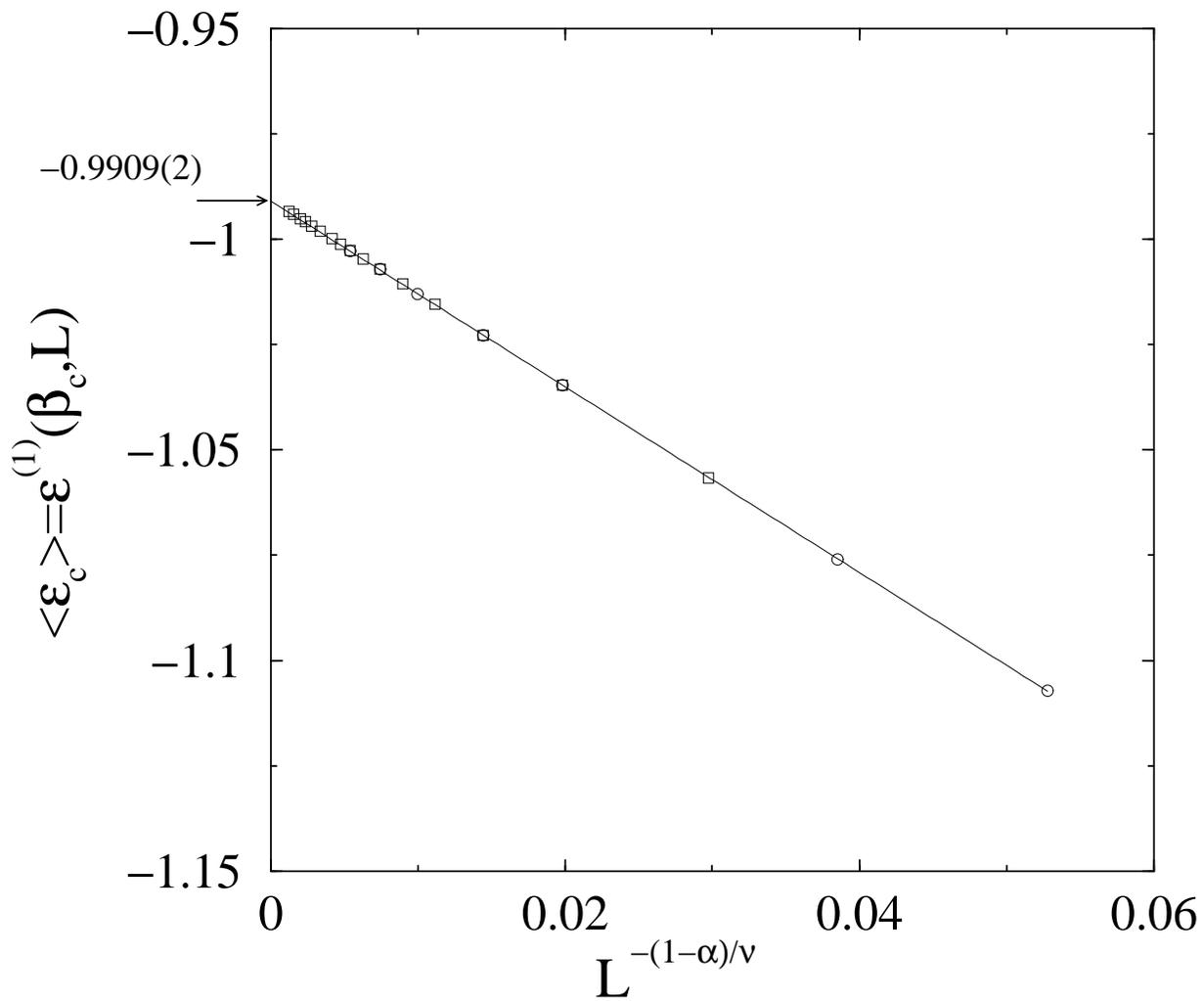}
\caption{
The canonical mean 
of the energy density 
for the critical d=3 Ising model as a
function of system size \protect\cite{unitsofenergy}. 
The statistical errors are an order of magnitude smaller than the system size.
The points marked $\diamond$ are taken from reference
\protect\cite{hasenbusch}.
The relevant parameters have been assigned the values
\protect\cite{blote}:
$\alpha=0.108$, $\nu=0.63067$ and $\beta _c=0.2216544$. 
The arrow identifies the
best-fit value for the intercept, prescribing the constant $\epsilon_c$
(Eq.~\ref{eq:meanen}).}
\end{figure}

\begin{figure}[h]
\epsfxsize=160mm \epsffile{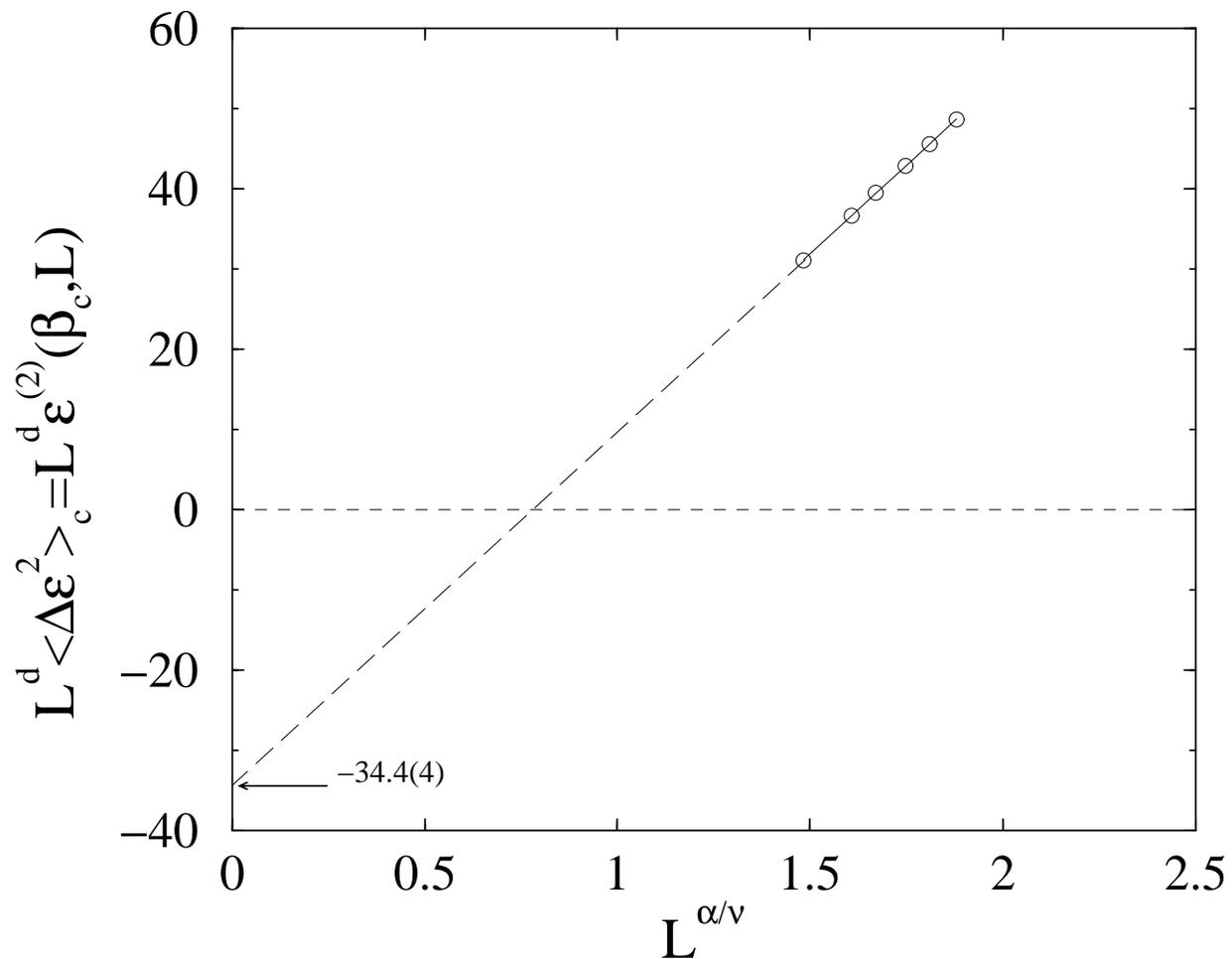}
\caption{
The canonical variance 
of the energy density for the critical
d=3 Ising model as a
function of system size \protect\cite{unitsofenergy}. 
The statistical errors are an order of magnitude smaller than the system size.
The parameters are as specified in Fig. 1.
The arrow identifies the best fit value for the intercept, prescribing the
constant $c_{0c}$ (Eq.~\ref{eq:varencor}).}
\end{figure}

\begin{figure}[h]
\epsfxsize=160mm \epsffile{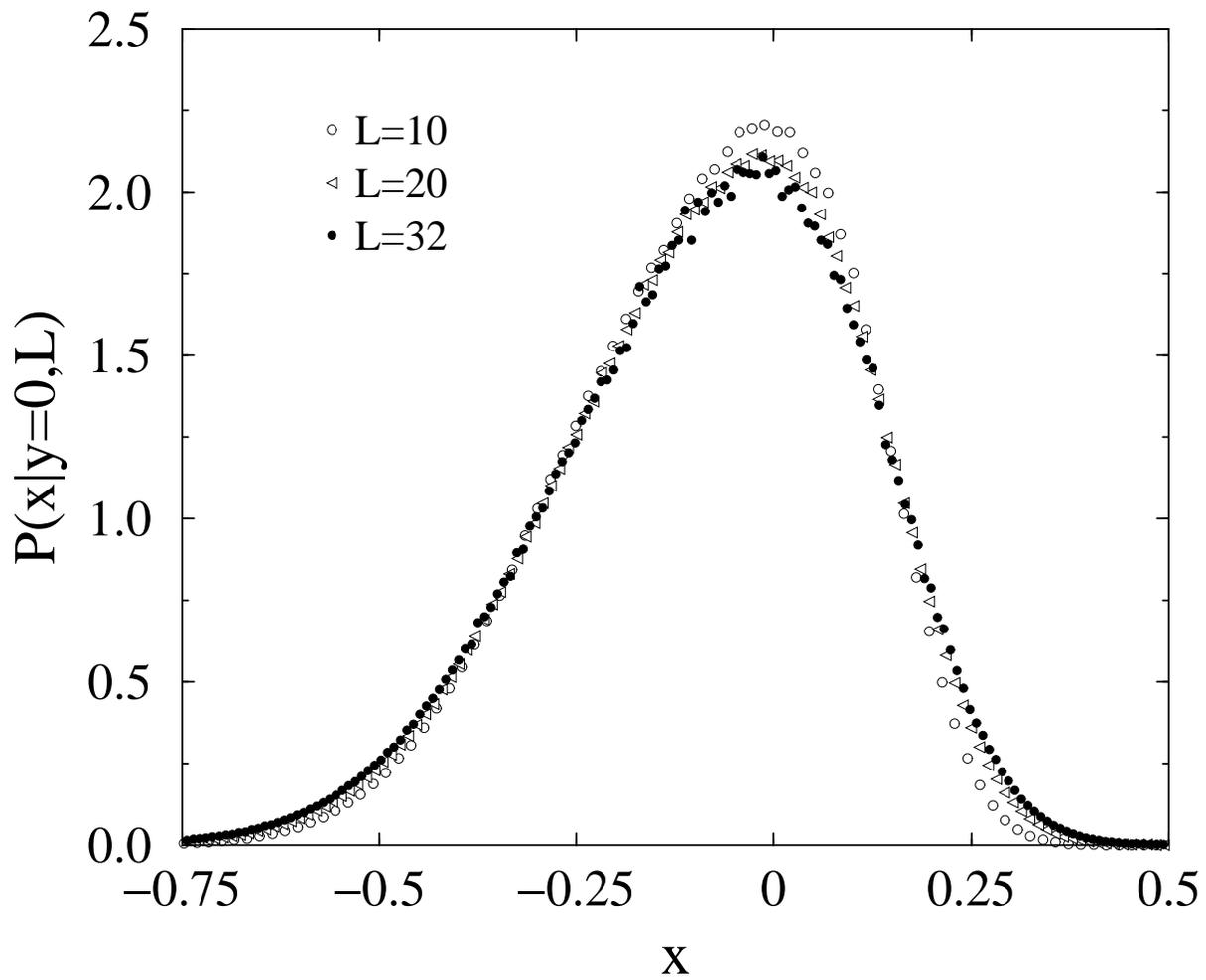}
\caption{
The canonical pdf (Eqs.~\ref{eq:canendistdef} - \ref{eq:canendistscalform}) 
of the scaled dimensionless energy density $x$ of the critical
d=3 Ising model, for a range of system sizes. The scaled variable is defined in
Eq.~(\ref{eq:xdef}) with the choice (cf Fig.~1) $\epsilon_c= -0.9909$. 
The scale factor $a_{\epsilon}$ implicit in the
scale of the $x$-variable is chosen such that
(cf Eq.~(\ref{eq:xdef})) $x= \protect\epsilon -\protect \epsilon_c$
for $L=10$.}
\end{figure}

\begin{figure}[h]
\epsfxsize=160mm \epsffile{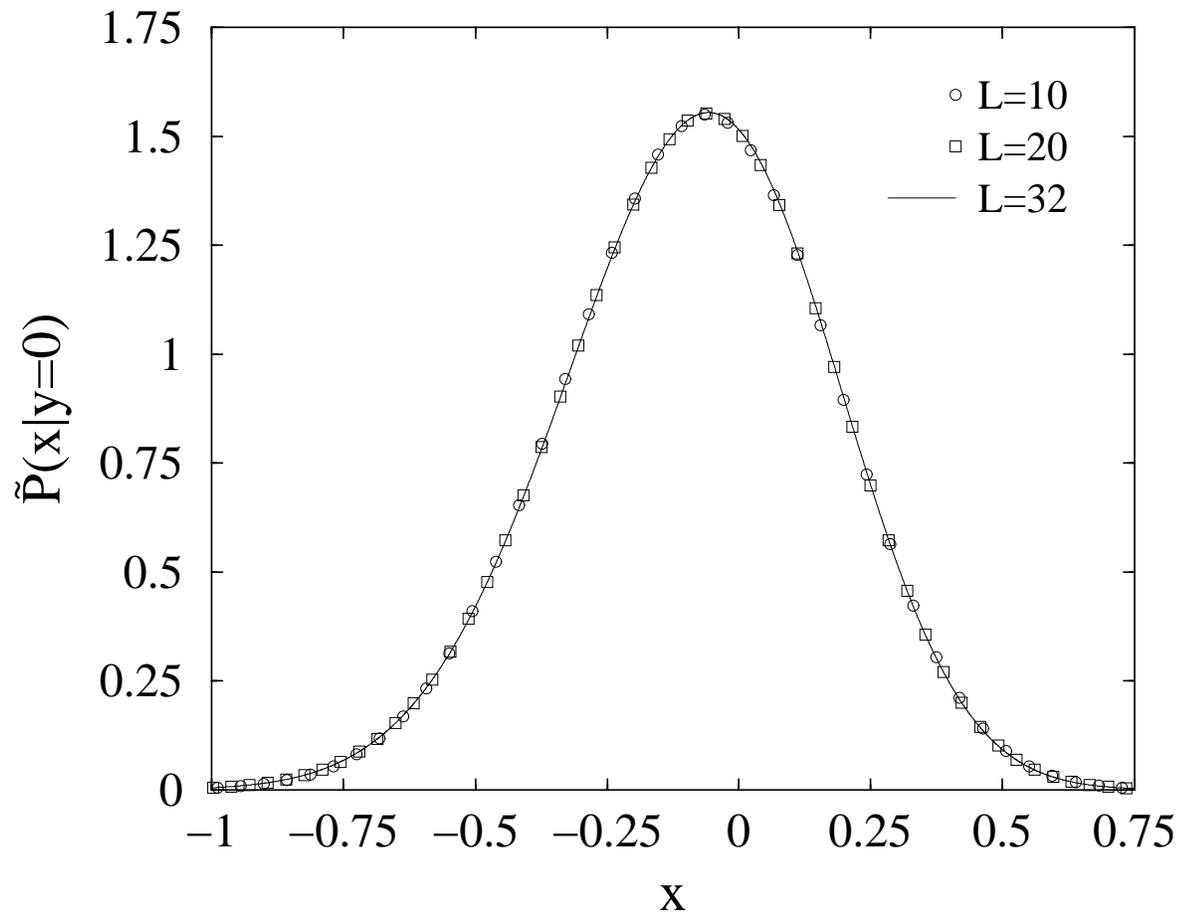}
\caption{
The data of Figure 3 with the effects of the 
non-scaling background convoluted out as prescribed by
Eq.~(\ref{eq:PtxyPxy}).
}
\end{figure}

\begin{figure}[h]
\epsfxsize=160mm \epsffile{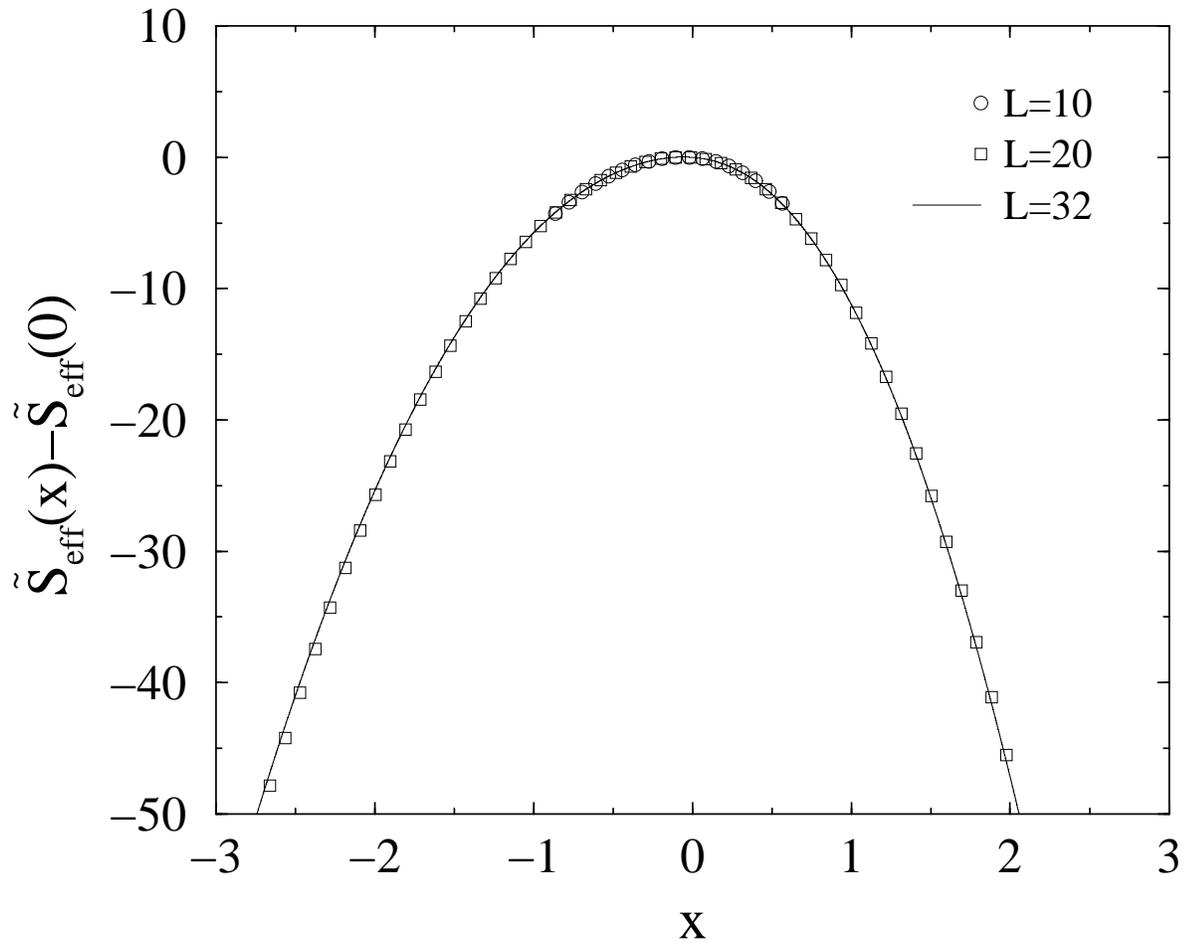}
\caption{
The finite-size  scaling function for the `effective' microcanonical entropy
$\tilde{\cal S}_{eff} (x)$ defined by Eq.~(\ref{eq:effSdef}) and deduced
from the critical canonical energy pdf, with the aid of
Eq.~(\ref{eq:effScalc}). Multi-histogram methods \protect\cite{multihist} have been used
to allow access to an extended range of $x$-values. 
}
\end{figure}

\begin{figure}[h]

\epsfxsize=160mm \epsffile{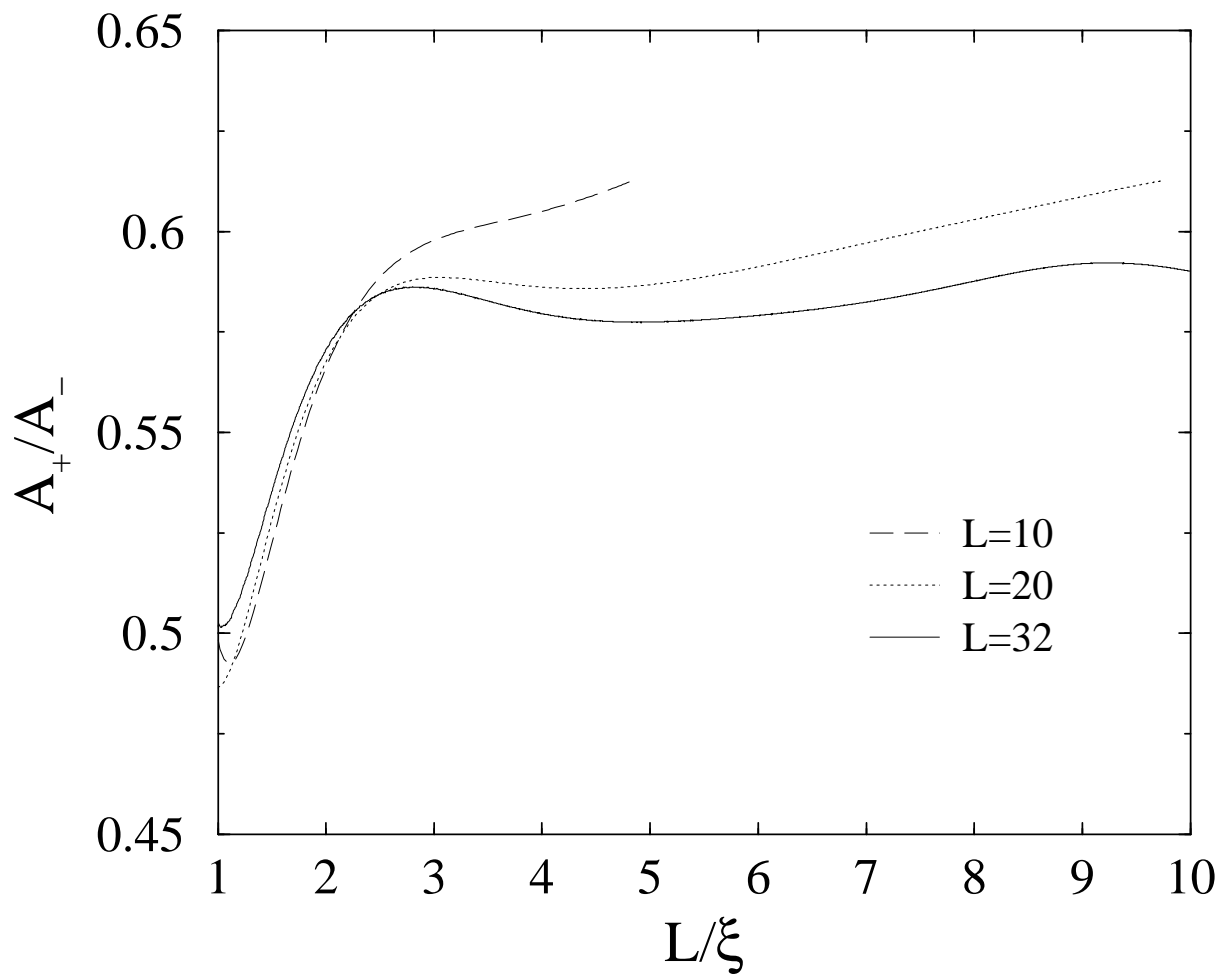}
\caption{Estimates of the  specific heat amplitude ratio, deduced from the decay
of the energy pdf (for different system sizes)  at `large' (positive and
negative) $x$-values. The estimates were determined by fitting to
pairs of ranges of $x$ values, with the  ranges forming each pair
being  centered on a common value of $L/\xi$, which forms the abscissa.
}

\end{figure}

\begin{figure}[h]
\epsfxsize=160mm \epsffile{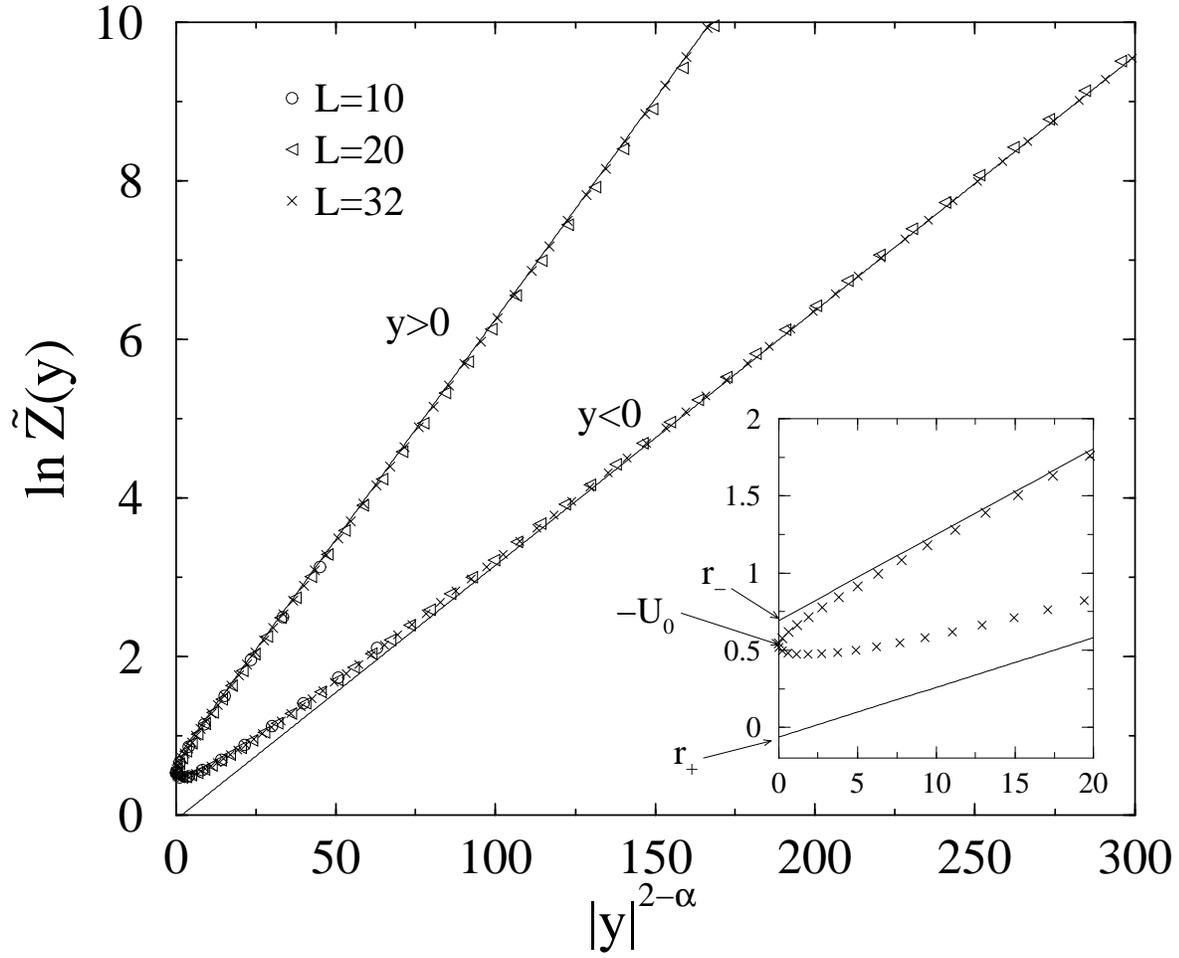}
\caption{
The function $\ln \tilde {\cal Z}(y)$ defined in Eq.~(\ref{eq:Zydef}), 
deduced from the microcanonical entropy 
(Fig. 5). The dimensionless variable $y$ 
provides a scaled representation
of the reduced (inverse) temperature (Eq.~\ref{eq:ydef}) .
The straight lines represent fits to the predicted asymptotic forms (Eq.~\ref{eq:largey}).
The arrows identify the
roles of the parameters $r_{\pm}$ (Eqs.~\ref{eq:anomalypmtwoa},b) and $U_0$
(Eq.~\ref{eq:anomalycritval}).
}
\end{figure}

\begin{figure}[h]
\epsfxsize=160mm \epsffile{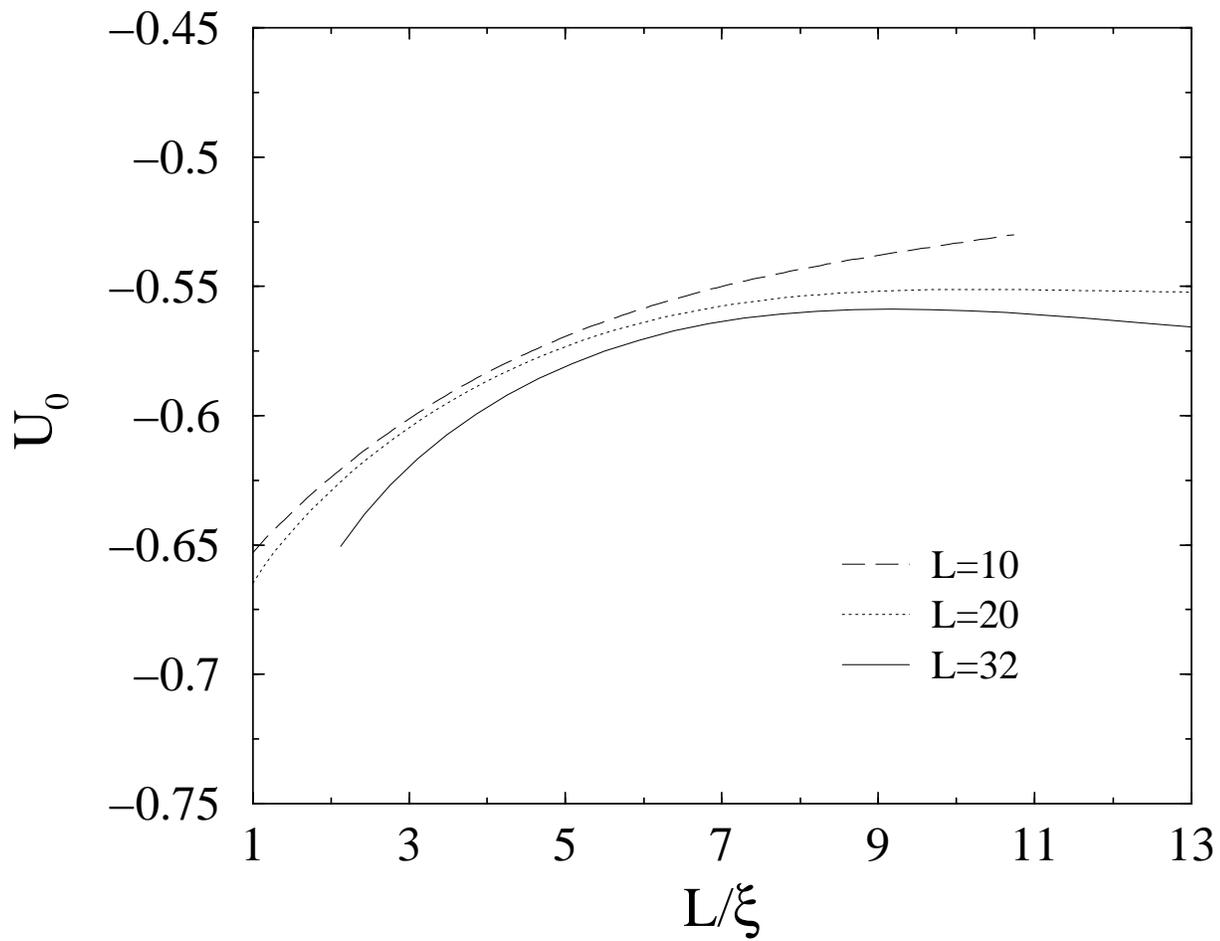}
\caption{Estimates of $U_0$ (Eq.~\ref{eq:anomalycritval})
determined, as in Fig. 6, for
a sequence of  different ranges of $L/\xi$ values; 
the mid-point of the range defines the
abscissa.         
}
\end{figure}

\begin{figure}[h]
\epsfxsize=160mm \epsffile{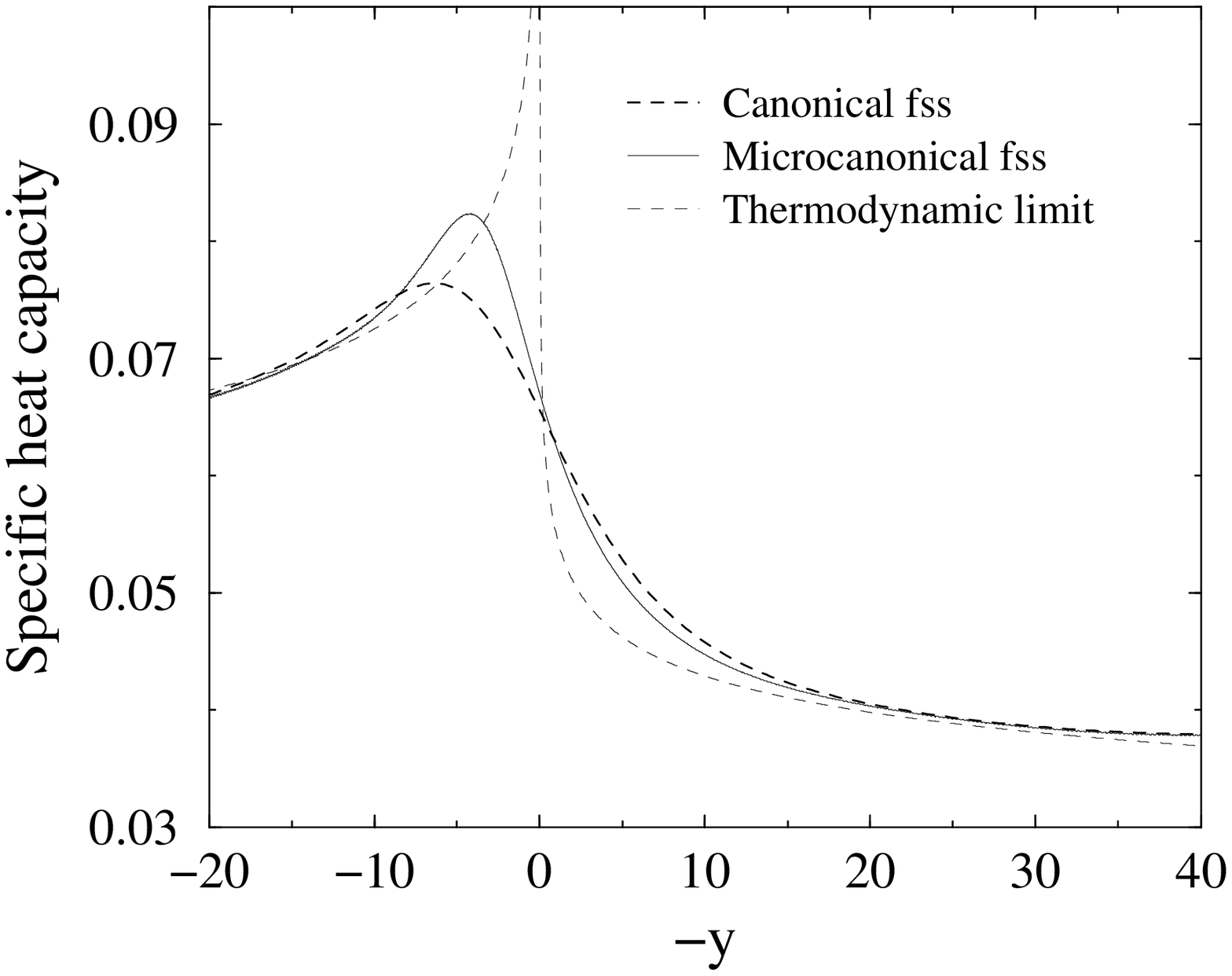}
\caption{
Comparison of the dimensionless microcanonical and canonical specific heat scaling
functions, $\tilde{c}^{\mu e}$
(Eq.~\ref{eq:microshscala})
and
$\tilde{c}^{ce}$
(Eq.~\ref{eq:canshscala}), plotted as a function of the scaled dimensionless
temperature (Eq~\ref{eq:ydef}).
The light dashed line shows
the power-law behavior characterizing the thermodynamic (large $\mid y \mid $) limit, extended back into
the finite-size-limited region.}
\end{figure}

\appendix

\renewcommand{\theequation}{\Alph{section}\arabic{equation}}

\section{Defining a density of states}

We discuss here, in general terms, the issues arising in {\em defining} a density
of states function for a system in which the energy spectrum is discrete. 
The conventional argument \cite{textbooks} makes the identification
\begin{equation}
\label{eq:convomegadef}
\Omega(\epsilon,L) \delta \epsilon = \sum_{E<E_s<E+\delta E} \Omega_s
\end{equation}
with the implicit assumption that the right hand side is
proportional to $\delta E$ ($\equiv L^d \delta \epsilon$). This requires
that:
\begin{enumerate}
\item[{\bf [C1]}] There exist many distinct levels $s$ within the interval $\delta E$.
\item[{\bf [C2]}] The level degeneracy $\Omega _s$ is slowly varying over the interval $E \rightarrow E +\delta E$.
\end{enumerate}
The fractional variation of $\Omega_s$ over the interval $\delta E$ can be estimated
using Eq.~(\ref{eq:microbetadef}); condition C2 can then be expressed in the form
\begin{equation}
\label{eq:omegavarn}
\frac{d\ln \Omega(\epsilon , L)}{d\epsilon}\delta \epsilon
= \beta^{\mu e}(\epsilon , L) \delta E \ll 1
\end{equation}
Taken together, conditions C1 and C2 thus
amount to the requirement that 
\begin{equation}
\label{eq:deltaEcond}
\delta E_I \ll \frac{1}{\beta^{\mu e}(\epsilon , L)} 
\end{equation}
where $\delta E_I$ characterizes the intrinsic discreteness of the energy
spectrum. This condition is trivially satisfied in the classical limit
(Appendix B1 considers one case explicitly). But there are obvious exceptions:
in the Ising model (Appendix B2) Eq.~(\ref{eq:deltaEcond}) is satisfied only at energies
corresponding to  microcanonical temperatures that are `high' on the scale of the
critical temperature. Or, to put it another way, $\Omega_s$ is certainly {\em
not} slowly varying over a range wide enough to embrace many system 
energy levels. We
must now recognize, however, that Eq.~(\ref{eq:convomegadef}) (along with its
implicit assumptions) does not faithfully reflect the conditions needed to
legitimize the transition from discrete (Eq.~\ref{eq:canonicalpfdisc}) to
continuum (Eq.~\ref{eq:canonicalpfdef}) representations. Instead of 
Eq.~(\ref{eq:convomegadef}) we require, rather, that we can consistently write
\begin{equation}
\label{eq:unconvomegadef}
\Omega(\epsilon,L) e^{-\beta E}\delta \epsilon = \sum_{E<E_s<E+\delta E} \Omega_s e^{-\beta E_s}
\end{equation}
where (while retaining condition C1) we must replace condition C2 by
\begin{enumerate}
\item[{\bf [C2A]}] $\Omega _s e^{-\beta E_s}$ is slowly varying over the
interval $E \rightarrow E +\delta E$, {\em if} that interval
lies in a range contributing significantly to the thermal properties at
temperature $\beta$.
\end{enumerate}
The range {\em contributing significantly \ldots} is centered on the saddle
point $\hat {E} = L^d \hat{\epsilon}$ (Eq.~\ref{eq:hatepsdef}). 
As a result, while condition C2 requires
Eq.~(\ref{eq:omegavarn}),
condition C2A requires only that
\begin{equation}
\label{eq:softomegavarn}
\frac{d^2\ln \Omega(\epsilon , L)}{d\epsilon ^2} \delta \epsilon ^2
= L^{-d} \left[ c^{\mu e}(\epsilon , L) \right ] ^{-1} \delta E^2  \ll 1
\end{equation}
where we have used Eq.~(\ref{eq:microsh}). Thus, in place of
Eq.~(\ref{eq:deltaEcond}), we need simply
\begin{equation}
\label{eq:softdeltaEcond}
\delta E_I \ll \left [L^d c^{\mu e}(\epsilon , L) \right ] ^{1/2} \simeq 
L^d\left[ \epsilon^{(2)}(\beta^{\mu e} , L) \right ] ^{1/2} 
\end{equation}
where the last step uses Eq.~(\ref{eq:cansh}), and $\beta ^{\mu e} = \beta^{\mu e}(\epsilon , L)$
(Eq.~\ref{eq:microbetadef}). This equation expresses more explicitly the
implications of condition C2A. A density of states function will exist in the
operational sense (Eq.~\ref{eq:canonicalpfdef}) that it may be used to compute
thermal properties at a given temperature  as long as the canonical energy
distribution (for that temperature) is broad on the scale of the
intrinsic discreteness of the energy spectrum.

\section{Density of states of simple models}

Here we show that the general form for the density of states function proposed
in Eq.~(\ref{eq:omegaansatz}) is consistent with exact results for two simple
models.

\subsection{Quasi-continuous energy spectrum: harmonic lattice model}

Consider a system (a harmonic model of  the vibrations of a crystal structure,
for example)  whose energy spectrum is that of $N$ weakly-interacting harmonic
oscillators, with associated frequencies $\nu_j, j=1\ldots N$.
Then
\[
E(\{n\}) = h \sum_{j=1}^{N} n_j\nu_j \equiv \sum_{j=1}^{N} \epsilon_j
\]
gives the energy  of a microstate in which (for each $j$)  mode $j$ 
has quantum number $n_j$. We consider the classical ($h \rightarrow 0$) limit,
in which the energy levels are quasicontinuous.
In this case $\delta E_I \sim h\nu_{\em min}$,  Eq.~(\ref{eq:deltaEcond})
is satisfied, and
we may proceed as in Eq.~(\ref{eq:convomegadef}) to write 
\[
\Omega (\epsilon , N)  = \frac{1}{\delta \epsilon } 
\sum_{\{n\}} D(E(\{n\}) -N\epsilon)
\]
where 
\[
D(X)= \left\{ \begin{array}{ll} 1 & \hspace*{0.2cm}\mbox{if }0< X < N\delta \epsilon \\
0& \hspace*{0.2cm}\mbox{otherwise}
\end{array}
\right .
\]
while $\epsilon\equiv E/N$ is the energy per oscillator.
In the $h \rightarrow 0$ limit the sums on $n_j$ can be replaced 
by integrals on $\epsilon_j$ to give
\[
\Omega (\epsilon , N) =
\epsilon^{-1} E^{N} Q(N) I(N)
\]
where
\[
I(N)=  \prod_{j=1}^{N} \int_0^1 dx _j \delta(1- \sum_j x_j )
\]
and
\[
Q(N) =  \prod_{j=1}^N\frac{1}{h \nu _j}
\]
Writing an integral representation of the $\delta$-function we find \cite{gradshteyn}
\[
I(N)=  \frac{1}{2\pi} \int_{-\infty}^{+\infty} dh e^{-ih} \left [ \frac{e^{ih} -1}{ih}\right ]^N
= \frac{2}{\pi} \int_{0}^{\infty} dh cos \left[ (N-2) h \right] 
\left[ \frac{\sin h}{h} \right] ^N
=\frac{1}{\Gamma (N)}
\]
which may be approximated
using the asymptotic expansion for the $\Gamma$ function \cite{gradshteyn}
\begin{equation}
\label{eq:stirling}
\Gamma (z) =\sqrt{2\pi} z^{z-\frac{1}{2}} e^{-z} \left[1 +O(z^{-1}) \right ]  
\end{equation}

In analyzing the remaining (energy-independent) contribution  we suppose
that the frequency spectrum is that of a $d=1$ system of particles, with a gap. Then 
\[
\ln Q(N) =  -N \ln h - N q_N
\]
where the sum $q_N$ may be written in the form
\[
q_N = \frac{1}{N}\sum_{j=1}^N  \ln \nu_j =
\frac{1}{N} \sum _{r=0}^{N-1} H (\frac {2\pi r}{N})
\]
where $H(\theta)$ is periodic, and (invoking the assumed gap) $H(0)$ is non-zero.
It can be shown \cite{barberfisher}  that the $N\rightarrow \infty$ limit
of this sum
\[
q_\infty = \frac{1}{2\pi} \int_0^{2\pi} H(\theta) d\theta
\]
has finite-size corrections that are exponentially small in N.

Gathering these results together we conclude that the density of states has the
form of Eq.~(\ref{eq:omegaansatz}) with the identifications $N=L^d$ and
\begin{equation}
\label{eq:sdefsho}
s (\epsilon) =\ln \left[\frac{\epsilon}{h}\right] -q_{\infty} +1
\end{equation}

Appealing to Eq.~(\ref{eq:canonicalpfresone}) one can readily recover, as a check,
the canonical partition function
\[
Z(\beta, N) = (\beta h)^{-N} e^{-Nq_\infty}
\]

\subsection{Discrete energy spectrum: 1d Ising model}
Consider a $d=1$ Ising model of $N$ sites, with periodic boundary conditions. 
Choosing the ground state as the energy-zero, the energy density
for a macrostate of $M$ domain walls
is $\epsilon =M{\epsilon_I}/N$, where $\epsilon_I$ 
is the domain wall energy.
The number of microstates corresponding to macrostate $M,N$ is
\[
\Omega_{M,N} = \frac{2 \times N!}{(N-M)! M!}
\]
Appealing to  the asymptotic form (\ref{eq:stirling}) once more
we find that
\[
\Omega_{M,N} = 2 \sqrt{\frac{N}{2\pi}}\left [ x(1-x)\right]^{-1/2} x^{-Nx}
(1-x)^{-N(1-x)} \left[ 1+ O(N^{-1}) \right ]
\]
where $x\equiv M/N= \epsilon/\epsilon_I$.
In this case $\delta E _I = \epsilon_I$ and 
Eq.~(\ref{eq:deltaEcond}) is not in general satisfied. But, since
Eq.~(\ref{eq:softdeltaEcond}) is, we may still identify a density of states by
\[ 
\Omega(\epsilon, N) = \frac{1}{2\epsilon _I} \Omega_{M,N}
\]
which one  may then readily recast in the 
form of Eq.~(\ref{eq:omegaansatz}) with the identifications $N=L^d$ and
\begin{equation}
\label{eq:sdefising}
s(\epsilon) = -(1-x)\ln(1-x) -x\ln x 
\end{equation}
Again, as a check, one can readily use this result to recover 
the canonical free energy density in the form
\[
f(\beta, N) = -\frac{1}{N} \ln Z(\beta, N) = -\ln (2 \cosh K ) +K
\]
where $2K=\beta \epsilon _I$, and the last term reflects our choice of ground
state energy.

\section{A bound on the canonical specific heat}

We outline here an argument establishing that the maximum of the value of the
microcanonical specific heat provides an upper bound for the canonical specific
heat, within the asymptotic  scaling region. The argument {\em assumes} the 
concavity of the function $\tilde{\cal S}_{eff} (x)$; the concavity of $\tilde{\cal S} (x)$ is already
presupposed in the formulation of Eq.~(\ref{eq:omegaansatz}). 

We write the scaling function for the energy pdf
(Eq.~\ref{eq:canendistscalform}) in the form:
\begin{equation}
\tilde{P}(x| y) = Q(x,y) G(x-\hat{x}(y), - 1/\tilde{S}_{eff}''(x^{\star}) )
\label{eq:identity}
\end{equation}
where $G(z,b)$ is a gaussian of zero mean, and variance $b$; 
$\hat{x}(y)$ is the modal scaled energy, for a given $y$, the solution of
\begin{equation}
\frac{d\tilde{\cal S}_{eff} (x)}{dx} =y
\label{eq:xhatdef}
\end{equation}
and $x^{\star}$ locates the maximum of the microcanonical specific heat,
identified by the condition (Eq.~\ref{eq:microshscalb})
\begin{equation}
\tilde{S}_{eff}''(x) \le \tilde{S}_{eff}''(x^{\star})
\label{eq:xstardef}
\end{equation}
The function $Q(x,y)$ introduced in Eq.~(\ref{eq:identity})
is defined  by:
\setcounter{abc}{1}
\begin{equation}
Q(x,y) = Q_0 e^{T(x,y)}
\label{eq:Qdef}
\end{equation}
where
\addtocounter{abc}{1}
\addtocounter{equation}{-1}
\begin{equation}
T(x,y) = -xy +\tilde{\cal S}_{eff} (x) -\frac{\tilde{S}_{eff}''(x^{\star})}{2} (x-\hat{x}(y))^2 
\label{eq:Tdef}
\end{equation}
\setcounter{abc}{0}
while $Q_0$ is an $x$-independent constant, defined by normalization.
From the assumed concavity of $\tilde{\cal S}_{eff} (x)$ it is straightforward to show that,
for any given $y$, $T(x,y)$ is concave in $x$, with a single maximum at
$x=\hat{x}(y)$.

Now appealing to Eqs. (\ref{eq:identity}) and (\ref{eq:canshscalb})
we can write
\begin{eqnarray}
\tilde{c}^{ce}(y)& =&\tilde{x}^{(2)} (y)\\
&=& \int dx \tilde{P}(x| y) \left[ x-\tilde{x}^{(1)}(y) \right] ^2\nonumber\\
&\le& \int dx \tilde{P}(x| y) \left[ x-\hat{x}(y)\right] ^2\nonumber\\
&=&\int dz Q(\hat{x}(y) +z,y) G(z,-1/\tilde{S}_{eff}''(x^{\star})) z^2\nonumber\\
&=&\int dz \tilde{Q}(z,y) G(z,-1/\tilde{S}_{eff}''(x^{\star})) z^2
\label{eq:cceytoQ}
\end{eqnarray}
where
\begin{eqnarray}
\tilde{Q}(z,y) & =& \frac{1}{2} \left[Q(\hat{x}(y)+z,y) +Q(\hat{x}(y)-z,y) \right] \nonumber\\
 & =& \frac{Q_0}{2}
\left[e^{T(\hat{x}(y) +z,y)} +e^{T(\hat{x}(y) -z,y)}\right]
\label{eq:tQdef}
\end{eqnarray}

From the properties of the function $T(x,y)$ it is straightforward to 
show that $\tilde{Q} (z,y)$ has a single turning point (at  $z=0$), and that
there exists some $z_0(y)$ such that
\begin{equation}
\tilde{Q}(z,y) \left\{ 
\begin{array}{ll}
 >1& \mbox{ if } \mid z \mid < z_0 (y)\\
 <1& \mbox{ if }\mid z \mid > z_0 (y) 
\end{array}
\right.
\label{eq:z0def}
\end{equation}
Then, finally, appealing to 
Eqs. (\ref{eq:cceytoQ}) and (\ref{eq:microshscalb})
\begin{eqnarray}
 \tilde{c}^{ce}(y) - \tilde{c}^{\mu e}(x^{\star})& <& \int dz \left[\tilde{Q}(z,y) -1 \right ]
G(z,-1/\tilde{S}_{eff}''(x^{\star})) z^2\nonumber\\
& <& z_0^2 (y) \int dz \left[\tilde{Q}(z,y) -1 \right ]
G(z,-1/\tilde{S}_{eff}''(x^{\star})) \nonumber\\
& =&0
\end{eqnarray}
where the last step exploits normalization conditions.

It follows that the microcanonical specific heat maximum $\tilde{c}^{\mu e}(x^{\star})$ provides
an upper bound for the canonical specific heat.


\begin{references}
\bibitem{asympequals} We will use $\simeq$ to signify equality in such an
asymptotic sense.
\bibitem{negative} See reference~\cite{liufish}.  The results of that paper
 indicate that the range of reduced temperatures below criticality ($y>0$), in which the
 subdominant `background' contribution to the specific heat is effectively
 constant, is small.
\bibitem{unitsofenergy} The Ising system energy is measured in units of the coupling
constant $J$.
\bibitem{whyplusminus} The sign convention for the subscripts is chosen to match
  the conventions of critical phenomena, where amplitudes are labeled
  according to the sign of $t \sim T-T_c$ rather than $y \sim \beta -\beta _c$.
\bibitem{alsoliufish} The tuning of the choice of high-temperature 
series-expansion-variable in reference \cite{liufish} has qualitatively similar 
consequences for amplitude-ratio-determination.
\bibitem{givenalpha} Note that this assignment (including its uncertainty)
is conditional upon the choices of the other parameters specified in the caption
to Fig. 1.
\bibitem{resolution} 
Note that the sharpness of the microcanonical peak reported in Fig. 1
of \cite{huller} is also heavily dependent upon the 
parameters of the smoothing procedure used there: see \cite{naturalsmooth}.
\end{references}
\end{document}